\documentclass[12pt,english,letterpaper]{scrartcl}

\usepackage[OT1]{fontenc}
\usepackage[latin1]{inputenc}
\usepackage{amsmath}
\usepackage{graphicx}
\usepackage{babel}
\usepackage{amssymb}
\IfFileExists{url.sty}{\usepackage{url}}
                      {\newcommand{\url}{\texttt}}

\makeatletter

\newif \ifpdf
\ifx \pdfoutput \undefined
      \pdffalse
   \else
      \pdftrue
\fi
\ifpdf
\usepackage{mathptmx}
\usepackage{helvet}
\usepackage{ae,aecompl}
\usepackage[pdftex,urlcolor=blue,colorlinks=true]{hyperref}
\else
\usepackage[ps2pdf,urlcolor=blue,colorlinks=true]{hyperref}
\fi
\begin{document}

\title{Bragg-induced orbital angular-momentum mixing in paraxial high-finesse
cavities}
\author{David H. Foster and \href{mailto:noeckel@uoregon.edu}{Jens U. N\"{o}ckel}\\
\medskip
Department of Physics, University of Oregon\\
1371 E 13th Avenue\\
Eugene, OR 97403\\
\url{http://darkwing.uoregon.edu/~noeckel}
}

\date{\emph{To be published}, Optics Letters \textbf{29}, December 1, 2004}

\maketitle

%\affiliation{Department of Physics, University of Oregon, 1371 E 13th Avenue,
%Eugene, OR 97403}

%\date{\emph{to be published}, Optics Letters \textbf{29}, December 1, 2004}

%\keywords{OCIS codes: 230.5750, 230.5440, 230.3990, 260.5430}

\begin{abstract}
Numerical calculation of vector electromagnetic modes of plano-concave
microcavities reveals that the polarization-dependent reflectivity
of a flat Bragg mirror can lead to unexpected cavity field distributions
for nominally paraxial modes. Even in a rotationally symmetric resonator,
certain pairs of orbital angular momenta are necessarily mixed in
an excitation-independent way to form doublets. A characteristic mixing
angle is identified, which even in the paraxial limit can be designed
to have large values. This correction to Gaussian theory is zeroth-order
in deviations from paraxiality. We discuss the resulting nonuniform
polarization fields. Observation will require small cavities with
sufficiently high Q. Possible applications are proposed. 
\end{abstract}
\maketitle
\noindent Unconventional beam profiles in free-space paraxial optics
have recently received renewed attention, based in large part on the
new degrees of freedom made accessible in light-matter interactions
when Gaussian beams carry not only spin but orbital angular momentum
\cite{PadgettPhysicsToday,GrierTweezers}. To realize the full potential
of the resulting rich phenomenology in quantum optical applications
\cite{milburn}, it is an important next step to turn from free-space
optics to ultrahigh-Q resonators. In this Letter, we study numerically
a type of cavity that enables robust and spectrally addressable creation
of \emph{nearly paraxial} light fields with orbital and polarization
profiles that are, surprisingly, \emph{not} predicted by the standard
solutions of paraxial theory. We build upon the initial examination
in \cite{methods, SPIE}.

\begin{figure}[!hbt]
\centering{\includegraphics[%
  width=0.75\columnwidth]{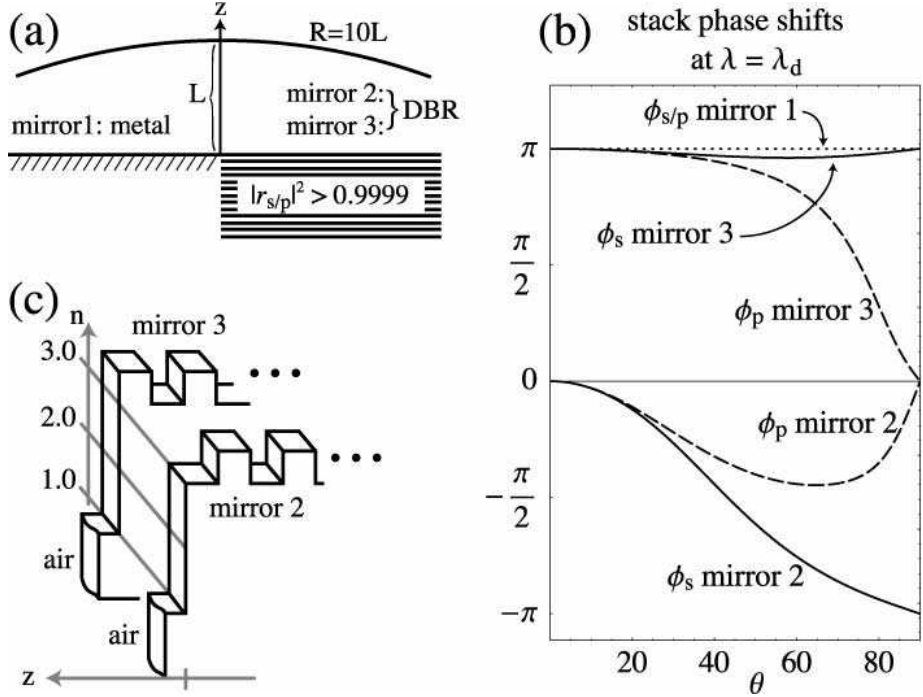}}

\caption{\label{Fig1}(a) Schematic cross section of the cavity. The dome
is metal, the base is either metal or Bragg mirror. The radius of
curvature $R$ is ten times larger than the length $L$. (b) Polarization-dependent
reflectivity phases \protect{}$\phi_{\text{s/p}}\equiv\arg(r_{\text{s/p}})$
of the different mirrors used in the calculation, versus angle of
incidence $\theta$. (c) Index profiles of the quarter-wave Bragg
mirrors used in Figs. 2 and 3. Each has 36 pairs of layers of index
3.0 and 3.5.}
\end{figure}

Beams carrying orbital angular momentum require an axially symmetric
environment, and hence the cavity in this study has a rotation axis
$z$. The dome-shaped resonator is shown in Fig. \ref{Fig1} (a).
Standard paraxial wave solutions with definite orbital angular momentum
are the \emph{Laguerre-Gauss beams} (LG). They are labeled by the
number $p$ of radial nodes, and the orbital angular momentum quantum
number $\ell$: in polar coordinates $\rho,\phi$, \begin{equation}
\mathrm{L}\mathrm{G}_{p}^{\ell}(\rho,\phi)\propto\left(\frac{\sqrt{2}\rho}{w}\right)^{|\ell|}\mathrm{L}_{p}^{|\ell|}\left(\frac{2\rho^{2}}{w^{2}}\right)e^{-\rho^{2}/w^{2}}e^{i\ell\phi},\label{eq:LGDef}\end{equation}
 where $\mathrm{L}_{p}^{|\ell|}(x)$ is an associated Laguerre polynomial
and $w$ is the beam waist radius. All beams of the same \emph{order}
$N\equiv2p+|\ell|$ form a degenerate manifold in the sense that their
longitudinal wave functions are the same for a given frequency $\omega$.
The degree of degeneracy is $g=2(N+1)$, including two possible polarization
states (e.g. right and left circular, with Jones vectors $\hat{\sigma}^{\pm}$);
this allows the formation of linear combinations of Laguerre-Gaussians
to obtain a wide variety of transverse beam profiles. The cross-sectional
electric field within a degenerate LG manifold of order $N$ can be
expanded as\begin{equation}
\left(\begin{array}{c}
E_{x}\\
E_{y}\end{array}\right)=\sum_{p,\ell}\mathrm{L}\mathrm{G}_{p}^{\ell}(\rho,\phi)\left[A_{p,\ell}\hat{\sigma}^{+}+B_{p,\ell}\hat{\sigma}^{-}\right],\label{eq:Ndegenerate}\end{equation}
 where $\ell=-N+2j$ ($j=0,1,\ldots N$), and $p$ is fixed by $2p+|\ell|=N$.
This carries over to paraxial resonators where the discrete resonant
frequencies $\omega_{\nu,N}$ are labeled by a longitudinal index
$\nu$ and the mode order $N$, and do not depend on $\ell$ and $p$
individually. %

Our exact numerical solutions reveal that \emph{corrections} \cite{Babic}
to the standard paraxial resonator theory stated above lead to a \emph{splitting}
of the $N$-multiplets in (\ref{eq:Ndegenerate}). Compared to the
equidistant transverse mode spacing in $\omega_{\nu,N}$ (governed
by the Guoy phase shift), this additional non-paraxial splitting typically
occurs on a scale so fine that it can be disregarded in large resonators.
However, ultrahigh-finesse microcavities, which are now becoming technologically
feasible, will make it necessary to quantify this substructure. The
problem is then analogous to zeroth-order degenerate perturbation
theory, the small quantity being the \emph{paraxiality parameter},
$h\equiv\lambda/(\pi w)=\tan(\emph{divergence angle})$.

The question \emph{how} the degeneracy in (\ref{eq:Ndegenerate})
is lifted goes beyond paraxial theory. A first guess would be that
the new eigenmodes are the LG basis modes ${\textrm{LG}}_{p}^{\ell}\,\hat{\sigma}^{s}$
where $s=\pm1$ is the spin. All modes are then uniquely labeled by
$p$, $\ell$ and $s$. This is indeed what we find when modeling
the planar mirror in Fig.~\ref{Fig1} (a) as a perfect electric or
magnetic conductor ($E_{{\textrm{tangential}}}=0$ or $H_{\textrm{{tangential}}}=0$).

In this Letter, however, we focus attention on the case where the
planar mirror is a realistic Bragg stack. Then, \emph{most} of the
dome cavity modes are \emph{not} labeled by a unique $(p,\,\ell,\, s)$,
even for the smallest values of the paraxiality parameter $h$. What
makes distributed Bragg reflectors (DBR) different is their polarization-dependent
plane-wave reflectivity $r_{\textrm{{s/p}}}(\theta)$, where s/p denotes
linear polarization perpendicular/parallel to the plane of incidence
and $\theta$ is the angle of incidence. Figure \ref{Fig1} (b) illustrates
that reflection \emph{phase shifts} $\phi_{\textrm{{s/p}}}$ with
strong $\theta$ dependence may occur even if the DBR design wavelength
$\lambda_{\textrm{{d}}}$ is chosen so that the modes of interest
are at the center of the stop band, where $|r_{\textrm{{s/p}}}(\theta)|\approx1$.
We have calculated the fully vectorial cavity modes, using a recently
developed numerical technique combining a basis function expansion
in the dome region with the transfer matrix technique for the DBR
\cite{methods,SPIE}. What we describe below can be called DBR-induced,
paraxial spin-orbit coupling of light.

\begin{figure}[!hbt]
\centering{\includegraphics[%
  width=0.85\columnwidth]{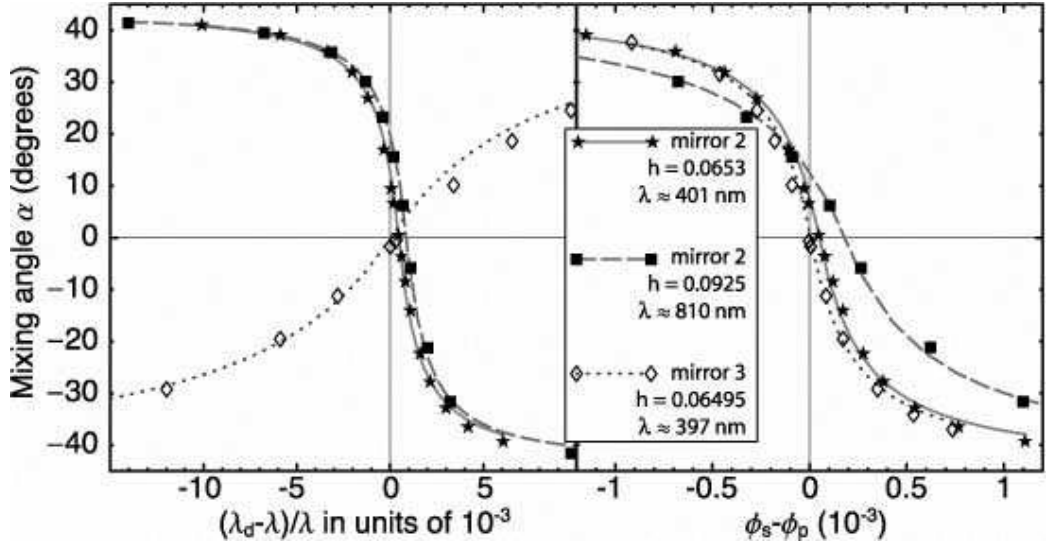}}

\caption{\label{cap:Fig2}Mixing angle $\alpha$ for numerically calculated
vector cavity modes, versus (a) relative detuning between the mean
wavelength $\lambda$ of the doublet $\mathbf{C}$, $\mathbf{D}$
and the Bragg design wavelength $\lambda_{\textrm{{d}}}$, and (b)
$\phi_{\text{s}}-\phi_{\text{p}}$. Line fits use the function given
in the text. Cavity dimensions are $L=10\mu$m and $R=100\mu$m. }
\end{figure}

The numerical method is not restricted to near-paraxial modes\cite{methods},
and the results presented here are observed over a wide range of $h$;
but we shall focus on the limit $h\ll1$ where $\{{\textrm{LG}}_{p}^{\ell}\,\hat{\sigma}^{s}\}$
should constitute a suitable basis in which to express the (transverse)
cavity fields. The vectorial modes can be chosen to be eigenfunctions
of the \emph{total angular momentum} around the $z$ axis, with integer
eigenvalues $m$. If orbital and spin angular momenta $\ell$ and
$s$ are {}``good quantum numbers'', then $m=\ell+s$. As illustrated
in Table 1 of Ref.~\cite{methods}, specifying $N$ and $m$
in (\ref{eq:Ndegenerate}) singles out \emph{pairs} of $\mathrm{LG}_{p}^{\ell}\,\hat{\sigma}^{s}$,
with $\ell=m\pm1$ (unless $|m|=N+1$, in which case only $|\ell|=N$
occurs). We shall call these pairs \emph{doublets} because their paraxial
degeneracy is in fact lifted. This is to be distinguished from a remaining
reflection-induced \emph{exact} degeneracy between modes differing
only in the sign of $m$. By fixing $m$, the latter degeneracy can
be disregarded for our purposes.

The lowest order for which the above doublets exist is $N=2$, and
we consider this case from now on. Both $\mathbf{A}\equiv\mathrm{LG}_{1}^{0}\hat{\sigma}^{+}$
and $\mathbf{B}\equiv\mathrm{LG}_{0}^{2}\hat{\sigma}^{-}$ are transverse
basis functions with $m=1$. If $m$ is a good quantum number but
$\ell$ and $s$ are not, then the transverse parts of the actual
$m=1$, $N=2$ cavity modes (denoted symbolically by $\mathbf{C}$,
$\mathbf{D}$) will be superpositions \begin{equation}
\left(\begin{array}{c}
\mathbf{C}\\
\mathbf{D}\end{array}\right)=\left[\begin{array}{cc}
\cos\alpha & -\sin\alpha\\
\sin\alpha & \cos\alpha\end{array}\right]\left(\begin{array}{c}
\mathbf{A}\\
\mathbf{B}\end{array}\right),\label{eq:CDRAB}\end{equation}
 where $\alpha$ is a \emph{mixing angle}: when $\alpha=0$, the modes
are well approximated by pure LG profiles. Among the different mirrors
shown in Fig.~\ref{Fig1}, only the conductor shows $\alpha=0$ in
general. The doublet's resonant frequencies satisfy $\omega_{\textrm{{D}}}-\omega_{\textrm{{C}}}\rightarrow0$
for $h\rightarrow0$, but the mixing angle $\alpha$ generally does
\emph{not} vanish in this paraxial limit.

\begin{figure}[!hbt]
\centering{\includegraphics[%
  width=1.0\columnwidth]{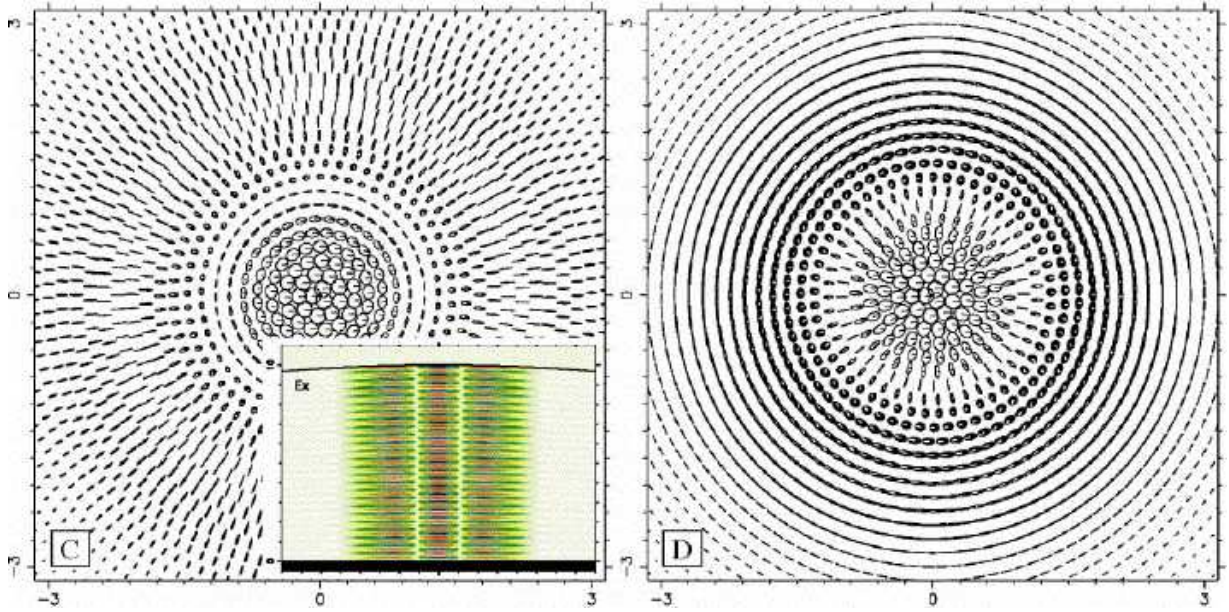}}

\caption{\label{cap:Fig3}Cross-sectional electric field of $\mathbf{C}$
and $\mathbf{D}$ for the point marked by the arrow in Fig.~\ref{cap:Fig2}
(a) (mirror design 2, $\lambda=401.1$nm, divergence angle $\theta\approx3.6^{\circ}$).
The ellipses represent only the $x$-$y$ components of the electric
field; white or black filling indicates sense of rotation. Each polarization
ellipse is decorated with a line pointing along the instantaneous
field vector to indicate the phase. Here the mixing angle is $\alpha=39^{\circ}$,
and $\omega_{\textrm{{D}}}-\omega_{\textrm{{C}}}=2\pi\times3.4\textrm{{GHz}}$.
Inset: $E_{x}$ in the $x$-$z$ plane. }
\end{figure}

Figure \ref{cap:Fig2} (a) gives the variation of $\alpha$ as $\lambda_{\textrm{{d}}}$
is changed. The sigmoid shape of the curves means that in a typical
cavity, one will find the doublet $\mathbf{C}$, $\mathbf{D}$ at
a value of $|\alpha|$ near $45^{\circ}$. This is the \emph{furthest
from pure} LG \emph{modes} we can get, in spite of the rotational
symmetry of the whole cavity. Note that $\alpha=90^{\circ}$ again
describes pure LG states, and in going through a total variation by
$\Delta\alpha=90^{\circ}$ the modes exchange character. This dramatic
change occurs over a narrow $\lambda-\lambda_{\textrm{{d}}}$ interval,
going through the {}``unmixed'' state $\alpha=0$ at which point
the resonance frequencies $\omega_{\textrm{{C}}}$, $\omega_{\textrm{{D}}}$
exhibit an avoided crossing (not shown).

The non-conservation of orbital and spin angular momentum does not
rely on paraxiality: the $x$-$y$ field components of a cavity eigenstate
with total angular momentum $m$ are \emph{linear combinations} of
two opposing circular polarization fields which individually transform
under coordinate rotations according to \emph{different} orbital angular
momenta $\ell=m\pm1$; modes ${\textbf{A}}$ and ${\textbf{B}}$ in
(\ref{eq:CDRAB}) are the paraxial realization of this general
fact. A Bessel-wave expansion\cite{methods} of the mode shows that
one of the orbital wave functions can in fact be made to vanish exactly
\emph{if} $r_{\textrm{s}}(\theta)\equiv r_{\textrm{p}}(\theta)$ for
the planar mirror. This occurs for pure magnetic or electric conductors.
In the DBR cavity, on the other hand, we only have $r_{\textrm{s}}=r_{\textrm{p}}$
at $\theta=0$, cf. Fig.\ \ref{Fig1} (b). Since even the most paraxial
modes contain nonzero $\theta$ in their angular spectrum, one \emph{cannot}
generally factor the field into a fixed polarization vector multiplying
a scalar orbital wave function. Within the conventional paraxial approach,
where this factorization is performed at the outset, the consequences
of a finite $\alpha$ at nearly vanishing $h$ are lost.

By changing the detuning of the doublet from $\lambda_{\textrm{{d}}}$,
we also change the phase difference $\phi_{\text{s}}-\phi_{\text{p}}$
at any finite angle of incidence $\theta$. As discussed elsewhere
\cite{Foster2}, one can identify an {}``effective'' ${\tilde{\theta}}\approx h$
at which to evaluate $\phi_{\text{s}}-\phi_{\text{p}}$. Plotting
$\alpha$ versus this variable $\phi_{\text{s}}-\phi_{\text{p}}$
in Fig. \ref{cap:Fig2} (b), the universality of the spin-orbit coupling
is revealed: the data differ widely in the mirror designs and wavelengths
used, but collapse onto very similar curves. The broad appearance
of the dotted curve in  \ref{cap:Fig2} (a) arises mainly because
the reflection phases of mirror 3 are less sensitive to detuning.
The data are fit by the function\cite{Foster2}, $\tan\alpha=-(\phi_{\text{s}}-\phi_{\text{p}}-\phi_{0})/\bigl(\Gamma+|\phi_{\text{s}}-\phi_{\text{p}}-\phi_{0}|\bigr)$.
The offset $\phi_{0}$ for attaining circular polarization ($\alpha=0$)
accounts for the fact that $\phi_{\text{s}}$ and $\phi_{\text{p}}$
are not equal at zero detuning when evaluated at finite $\theta$.
The crossover between $\alpha\approx\pm45^{\circ}$ has width $\Gamma$,
and persists in the zeroth-order limit $h\rightarrow0$ where $\Gamma\rightarrow0$.

In Figure \ref{cap:Fig3}, we illustrate the polarization patterns
produced by the spin-orbit coupling. Polarization ellipticity and
orientation vary spatially. In particular, the single mode ${\textbf{C}}$
exhibits circular polarization near the axis but radial linear polarization
within its second radial maximum; in mode ${\textbf{D}}$, a crossover
from circular to azimuthal linear polarization occurs. The $L=10\mu$m
cavity height in this example allows $\omega_{\textrm{{D}}}-\omega_{\textrm{{C}}}$
to be resolved despite the finite widths of the modes, which is taken
into account in our numerics (cf.\ the DBR design in Fig.\ \ref{Fig1},
and the data on decay widths in Ref.\ \cite{methods}). Assuming
that both mirrors have a power reflectivity $r^{2}$ in the range
$(1-r^{2})<10^{-3.5\pm0.5}$, we estimate that the effects shown here
should be observable for cavities with paraxiality $0.09<h<0.2$ if the
height $L$ of the resonator lies in the window $12\lambda<L<60\lambda$.
The mode patterns in Fig. \ref{cap:Fig3} are governed by the mixing
angle $\alpha$; Fig.~\ref{cap:Fig2} suggests the intriguing possibility
of externally controlling $\alpha$ by changing the cavity dimensions
or Bragg wavelength $\lambda_{\textrm{d}}$.

When addressed by laser pulses at $\omega_{\textrm{{C}}}$ and $\omega_{\textrm{{D}}}$,
the doublet modes could interact with one, two, or more asymmetric
(polarization-sensitive) quantum dots embedded in the DBR. Thus the
mixed modes may allow a new scheme of quantum processing. Cavities
of the type studied here can also act as filters or laser emitters
of pure or mixed LG beams, useful as {}``optical tweezers''\cite{GrierTweezers}
or as carriers of information\cite{Galvez}.

Provided that at least one of the mirrors is a Bragg stack, \emph{any
stable cavity should exhibit paraxial mixing of orbital angular momenta}.
Its observability hinges on the ability to distinguish true degeneracies
from {}``quasi''-degeneracies caused by the breakdown of Eq.\ (\ref{eq:Ndegenerate}).
Our quantitative analysis shows that the necessary requirements can
be met by realistic cavities that are being pursued for quantum-optics
applications.

We thank Mike Raymer for valuable discussions. This work was supported
by NSF Grant ECS-0239332.

\end{document}